\documentclass[11pt,a4paper]{article}
\pdfoutput=1
\usepackage{jheppub_mod}
\usepackage{rotating}
\usepackage{array}
\usepackage{amsmath}
\usepackage{amssymb}
\usepackage[normalem]{ulem}
\usepackage{slashed,hyperref}
\usepackage{gensymb}
\usepackage{cancel}
\title{
{\rm \vspace{-2cm}
{\normalsize
	\flushright TUM-HEP-1159/18,\,KIAS-P18091\\}
\vspace{0.6cm}}
Potential for probing three-body decays of Long-Lived Particles with MATHUSLA
}

\author[a,b]{Alejandro Ibarra,}
\author[c]{Emiliano Molinaro}
\author[d]{and Stefan Vogl}

\affiliation[a]{Physik-Department T30d, Technische Universit\"at M\"unchen, James-Franck-Stra\ss e, 85748
Garching, Germany}
\affiliation[b]{School of Physics, Korea Institute for Advanced Study, Seoul 02455, South Korea}
\affiliation[c]{Department of Physics and Astronomy, University of Aarhus,
Ny Munkegade 120, DK-8000 Aarhus C, Denmark}
\affiliation[d]{Max-Planck-Institut f\"ur Kernphysik,
Saupfercheckweg 1, 69117 Heidelberg, Germany}

\emailAdd{ibarra@tum.de}
\emailAdd{molinaro@phys.au.dk}
\emailAdd{stefan.vogl@mpi-hd.mpg.de}

\abstract{
	Several extensions of the Standard Model predict the existence of Long-Lived Neutral Particles (LLNPs) with masses in the multi-GeV range and decay lengths of ${\cal O}(100\,{\rm m})$ or longer. These particles could be copiously produced at the LHC, but the decay products cannot be detected with the ATLAS or CMS detectors. MATHUSLA is a proposed large-volume surface detector installed near ATLAS or CMS aimed to probe scenarios with LLNPs which offers good prospects for disentangling the physics underlying two-body decays into visible particles. In this work we focus on LLNP decays into three particles with one of them being invisible, which are relevant for scenarios with low scale supersymmetry breaking, feebly interacting dark matter or sterile neutrinos, among others. We analyze the MATHUSLA prospects to discriminate  between two- and three-body LLNP decays, as well as the prospects for reconstructing the underlying model parameters.	
}

\begin{document}	
\maketitle

\section{Introduction}
\label{sec:intro}

Some well motivated scenarios of Physics Beyond the Standard Model (BSM) predict the existence of Long-Lived Neutral Particles (LLNPs) with masses in the multi-GeV range and decay lengths longer than the size of the ATLAS or CMS detectors. If the lifetime is longer than a few minutes, the late particle decays could have a significant impact on the abundances of primordial elements \cite{Kawasaki:2004qu,Jedamzik:2006xz} or the shape of the energy spectrum of the cosmic microwave background radiation~\cite{Hu:1993gc,Ellis:1990nb}. Conversely, the non-observation of statistically significant differences in the data with respect to the predictions of the Standard hot Big Bang scenario sets stringent limits on various BSM frameworks.  Unfortunately, the current LHC detectors have a limited sensitivity to scenarios with particles with a decay length between  $\sim 10$ m and $\sim 10^7$ m.

Recently, a number of experiments with an enhanced sensitivity to long-lived particles have been proposed, {\it i.e.} MATHUSLA~\cite{Chou:2016lxi}, FASER~\cite{Feng:2017uoz} or CODEX-b~\cite{Gligorov:2017nwh}.  In particular the MATHUSLA proposal aims for a high sensitivity to LLNPs decaying into two Standard Model particles with a decay length of $\mathcal{O}(100)\;\mbox{m}$~\cite{Chou:2016lxi}, which is of great relevance for a large range of new physics scenarios, notably scenarios with an exotic Higgs~\cite{Curtin:2018mvb}. 

In this work we explore the capability of MATHUSLA to identify and study LLNP decays into three particles, more specifically when one of these is invisible, such that the final state contains only two visible particles.  This decay topology is realized in frameworks where the Standard Model is extended with a BSM sector,  charged under a new unbroken or mildly broken symmetry, and where the lightest BSM particle is a dark matter candidate and the next-to-lightest BSM particle has sizable interactions with the Standard Model. In this case, the next-to-lightest BSM particle could be copiously produced in collider experiments and would generically decay into Standard Model particles and the dark matter candidate.  Concrete examples of this kind of models are supersymmetric frameworks with R-parity conservation where the lightest neutralino decays into the gravitino and a fermion-antifermion pair \cite{Dimopoulos:1996vz,Bagger:1996bt,Ambrosanio:1996jn,Dimopoulos:1996va},  or certain models of feebly interacting massive particle (FIMP) as dark matter \cite{Hessler:2016kwm, Arcadi:2013aba,Calibbi:2018fqf}. Alternatively, the signature can also arise in supersymmetry with R-parity violation where the lightest neutralino decays into a neutrino and a fermion-antifermion pair~\cite{Meade:2010ji,Graham:2012th} or  in models with sterile neutrinos~\cite{Gorbunov:2007ak,Helo:2013esa,Accomando:2016rpc,Jana:2018rdf,Deppisch:2018eth}. A classification of simplified models displaying displaced vertices and their signatures was recently presented in \cite{Buchmueller:2017uqu}.

\section{Two-body vs. three-body decays}
\label{sec:23}

The capability of MATHUSLA to probe scenarios with LLNPs is currently being explored~\cite{Curtin:2017izq,Curtin:2018mvb,Evans:2017lvd}. The experiment essentially consists of a large volume with a series of tracking layers on top. Its proposed location is on the surface, {\it i.e.} $\approx 100$ m above the experimental cavern, close to one of the LHC interaction points. In our analysis we adopt the benchmark proposal from \citep{Curtin:2017izq} which consists of a $200 \,\mbox{m} \times 200 \,\mbox{m} \times 25 \,\mbox{m}$ hall located $100 \,\mbox{m}$ upstream of the CMS or ATLAS detector. The detector features five layers of tracking material with  the first tracking layer at a height of $20$ m and the other four tracking layers placed above it with a separation of $1\,\mbox{m}$  from each other.   The spatial resolution of this setup is limited by the size of the pixels in the  tracking layer, which is currently projected to be $1 \, \mbox{cm}^2$. 
With this set-up, the experimental signal consists of $2\times 5$ hits in the tracking system, which allows to reconstruct the direction of the daughter particles and the location of the displaced vertex with an angular resolution of $\sim 2 \times 10^{-3}$ rad. It should be noted that this detector set-up does {\it  not} allow for the determination of the energy and/or momentum of the final state particles. Therefore, new approaches must be developed in order to disentangle the underlying Particle Physics model with the limited information provided by MATHUSLA.  

In this paper we will focus in the prospects of MATHUSLA to probe scenarios where the LLNP decays into three particles,  with one final state particle going undetected. As a first step it is crucial to determine whether the two observed tracks  are originating from a two-body decay, from a three-body decay, or from background. To this end, it is convenient to construct an observable quantity at MATHUSLA that can differentiate among them. 

Let $\vec P$ be the LLNP momentum  and $\vec p_i$ the momenta of the daughter particles. The directions of the particles are given by the normalized vectors, $\vec P/|\vec P|$, $\vec p_i/|\vec p_i|$.
For the two-body decay,  momentum conservation requires $\vec P=\vec p_1+\vec p_2 $.
Consequently, the triple product
\begin{equation}
	T=\frac{\vec P}{|\vec P|}\cdot\left( \frac{\vec p_1}{|\vec p_1|}\times \frac{\vec p_2}{|\vec p_2|}\right)
\end{equation}
is identically zero.  Strictly, the limited angular resolution of the experiment would lead to a non-vanishing value for $T$, but given the level of precision expected for MATHUSLA, {\it i.e.} an angular resolution of $\sim 0.002$ rad, the deviation of $T$ from zero is expected to be of this order\footnote{ In principle the interaction point is actually an interaction region and the position of the primary vertex is not known exactly. However, the size of the interaction region is described to excellent precision by a normal distribution with a standard deviation of $\approx 2$ cm~\cite{ATLAS-CONF-2010-027} in the beam direction, and less in the perpendicular direction. The angular uncertainty introduced by the position of the primary vertex can therefore be neglected.}.

Instead, for a three body decay $\vec P=\vec p_1+\vec p_2+\vec p_3$.
Correspondingly, one finds
\begin{equation}
T=\frac{\vec P}{|\vec P|}\cdot\left( \frac{\vec p_1}{|\vec p_1|}\times \frac{\vec p_2}{|\vec p_2|}\right)=\frac{\vec p_3}{|\vec P|}\cdot\left( \frac{\vec p_1}{|\vec p_1|}\times \frac{\vec p_2}{|\vec p_2|}\right) = \cos\theta \sin\phi\;,
\label{eq:T-3body}
\end{equation}
where $\theta$ denotes the angle between LLNP direction and the direction perpendicular to the decay plane spanned by the two tracks, while $\phi$ is the opening angle of the two tracks. Clearly, $T$ will be in general different from zero. Note also that $T$ is constructed from angular variables defined in the laboratory frame, which in turn depend on the angular distribution of the daughter particles in the rest frame (determined by the dynamics controlling the decay) and on the Lorentz factor of the decaying particle (controlled by the production mode at the LHC). In order to simplify our discussion we will assume in the following that the three-body LLNP decay is isotropic in the rest frame, and focus on the implications of different production modes.

To illustrate the impact of the LLNP production quantitatively we consider two representative benchmark scenarios.  In Scenario A the LLNP has mass $m_{\rm LLNP}<m_H/2$ and is produced in the decay of the Standard Model Higgs boson. In Scenario B the LLNP is the neutral component of a $SU(2)$ doublet, and is produced via the Drell-Yan process.~\footnote{	Scenario B can be in particular identified with a gauge mediated SUSY framework where the LLNP is a pure Higgsino, that decays into a gravitino and a lepton-antilepton pair with a small width suppressed by the scale of SUSY breaking.} The LLNP then decays into a lepton-antilepton pair and a light neutral stable particle (NSP) which is not detected. 
 We also assume for simplicity $m_{\rm NSP}\ll m_{\rm NNLP}$, motivated by supersymmetric scenarios with gravitino as lightest supersymmetric particle and gauge mediated supersymmetry breaking (such that the gravitino is predicted to be much lighter than the other sparticles), supersymmetric scenarios with neutralino as lightest supersymmetric particle and R-parity violation (where the neutralino decays into a fermion-antifermion pair and a neutrino), or sterile neutrinos (which decay into a fermion-antifermion pair and a neutrino). In these simplified scenarios the phenomenology in the MATHUSLA detector is completely determined by the LLNP mass. 

\begin{figure}[t]
	\centering
	\includegraphics[width=0.45\textwidth]{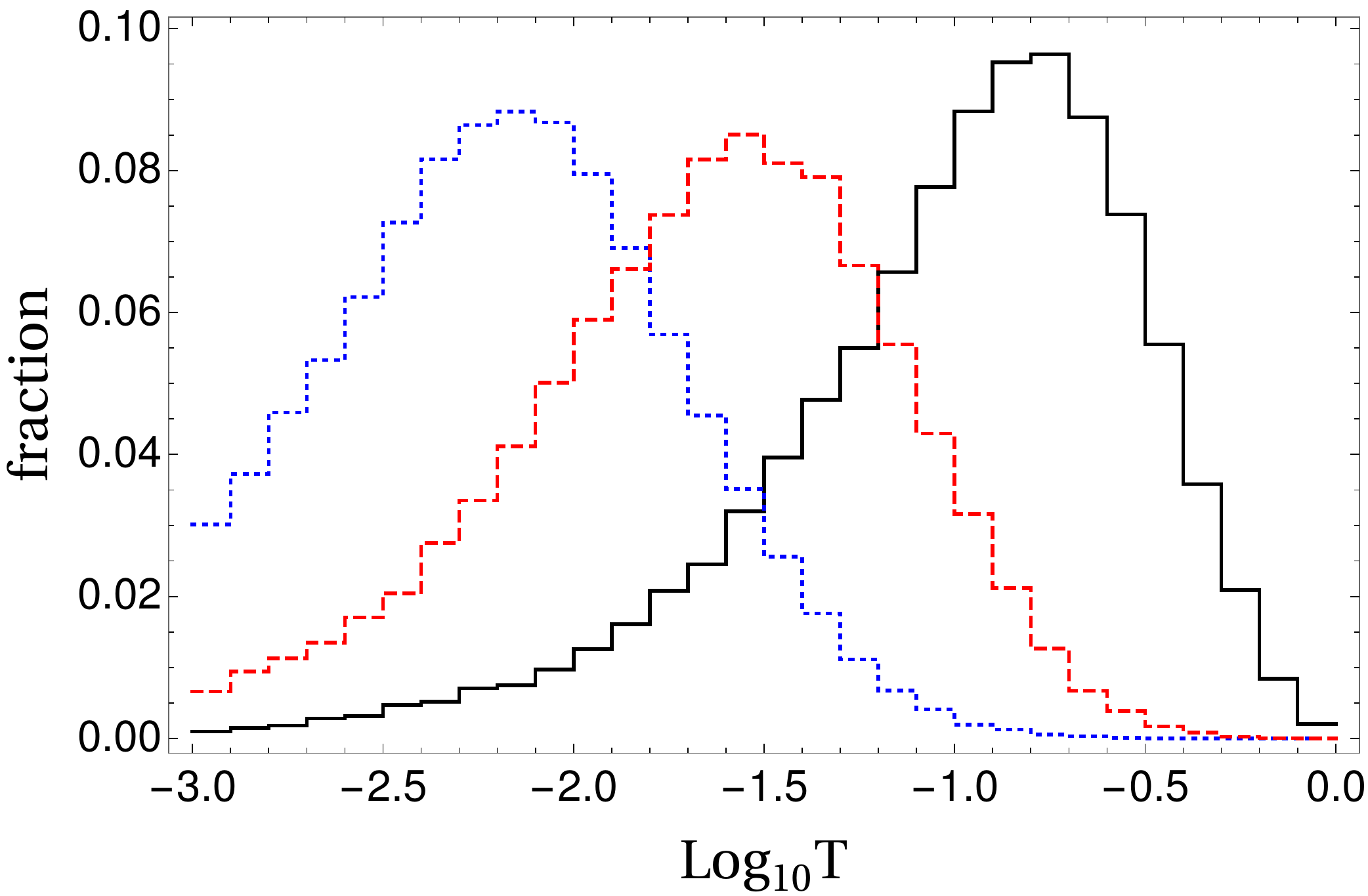}
	\includegraphics[width=0.45\textwidth]{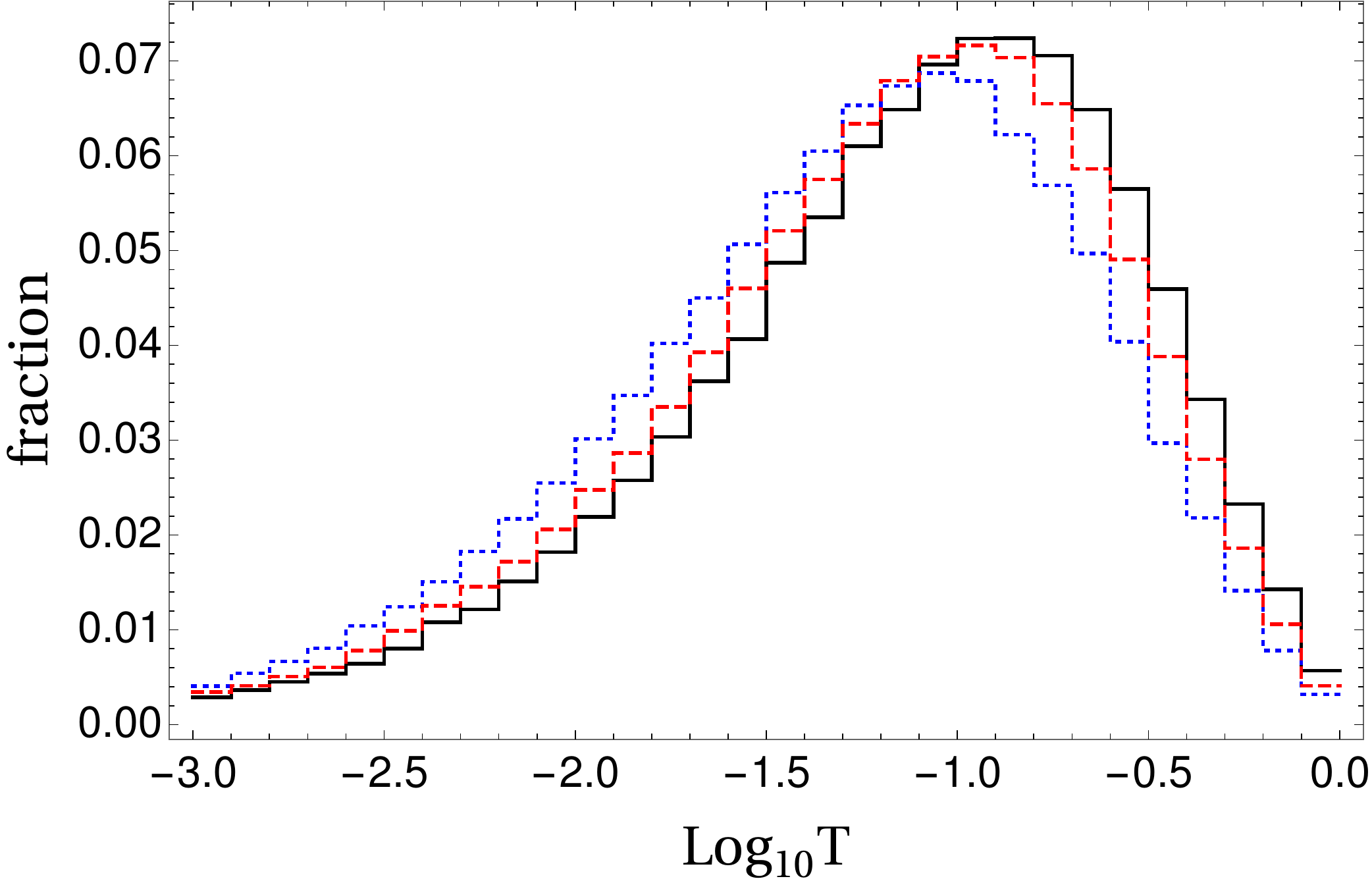}
	\caption{Distribution of events as a function of the triple product parameter $T$ assuming LLNP production via Standard Model Higgs decay (left plot) or via Drell-Yan (right plot). The black solid, red dashed and blue dotted lines correspond in the left plot to  $m_{\rm LLNP}=50$, $25$ and $12.5$ GeV respectively, and in the right plot to  $m_{\rm LLNP}=800$, $600$  and $400$ GeV.}
	\label{fig:Hist_T}
\end{figure}

We show in Fig.~\ref{fig:Hist_T} the distribution of  dilepton pair events as a function of the triple-product parameter $T$ for representative choices of the LLNP mass, $m_{\rm LLNP}$, assuming scenario A (left panel) or scenario B  (right plot). For scenario A, the typical LLNP Lorentz-factor is  $ m_H/ 2 m_{\rm LLNP}$. Consequently, the momentum and angular distributions in the laboratory frame (and accordingly the triple product parameter) depend strongly on $ m_{\rm LLNP}$, as apparent from Fig.~\ref{fig:Hist_T}. For the Drell-Yan process, on the other hand, the momentum distribution peaks just above threshold, and for $m_{\rm LLNP} \gg m_Z$ the slope of the high energy tail chiefly depends on the parton distribution functions at the relevant center of mass energy. This leads to a comparatively mild dependence of the $T$-parameter distribution on the LLNP mass.
It follows from the figure that in both scenarios, one generically expects a significant number of events with  $T \gtrsim 0.001$.
Therefore, and in view of the distinct $T$-distribution of the two- and three-body decays, one concludes that even a small sample of events would suffice to discriminate between these two possibilities.

\begin{figure}[t]
	\centering
	\includegraphics[width=0.49\textwidth]{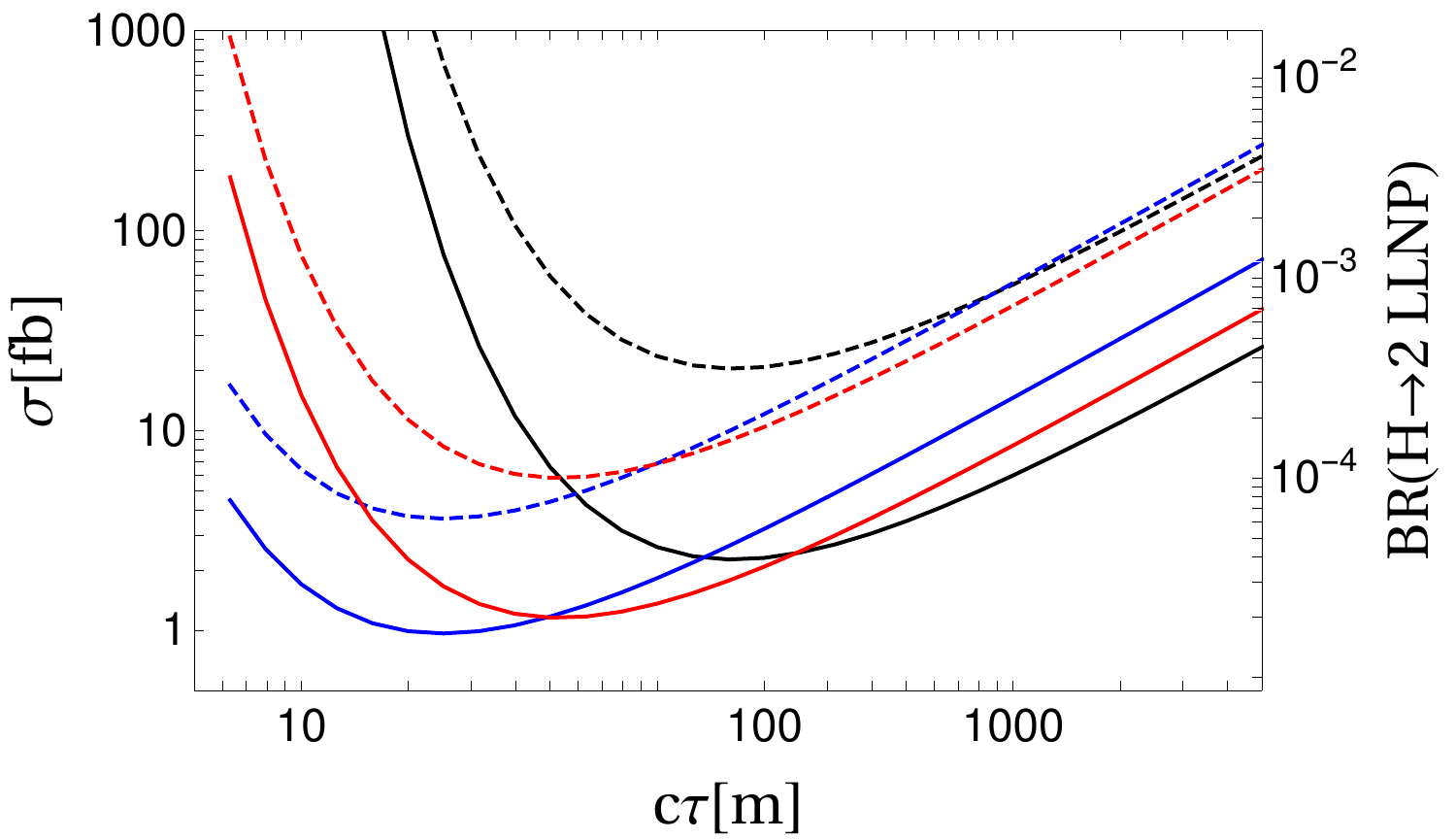}
	\includegraphics[width=0.49\textwidth]{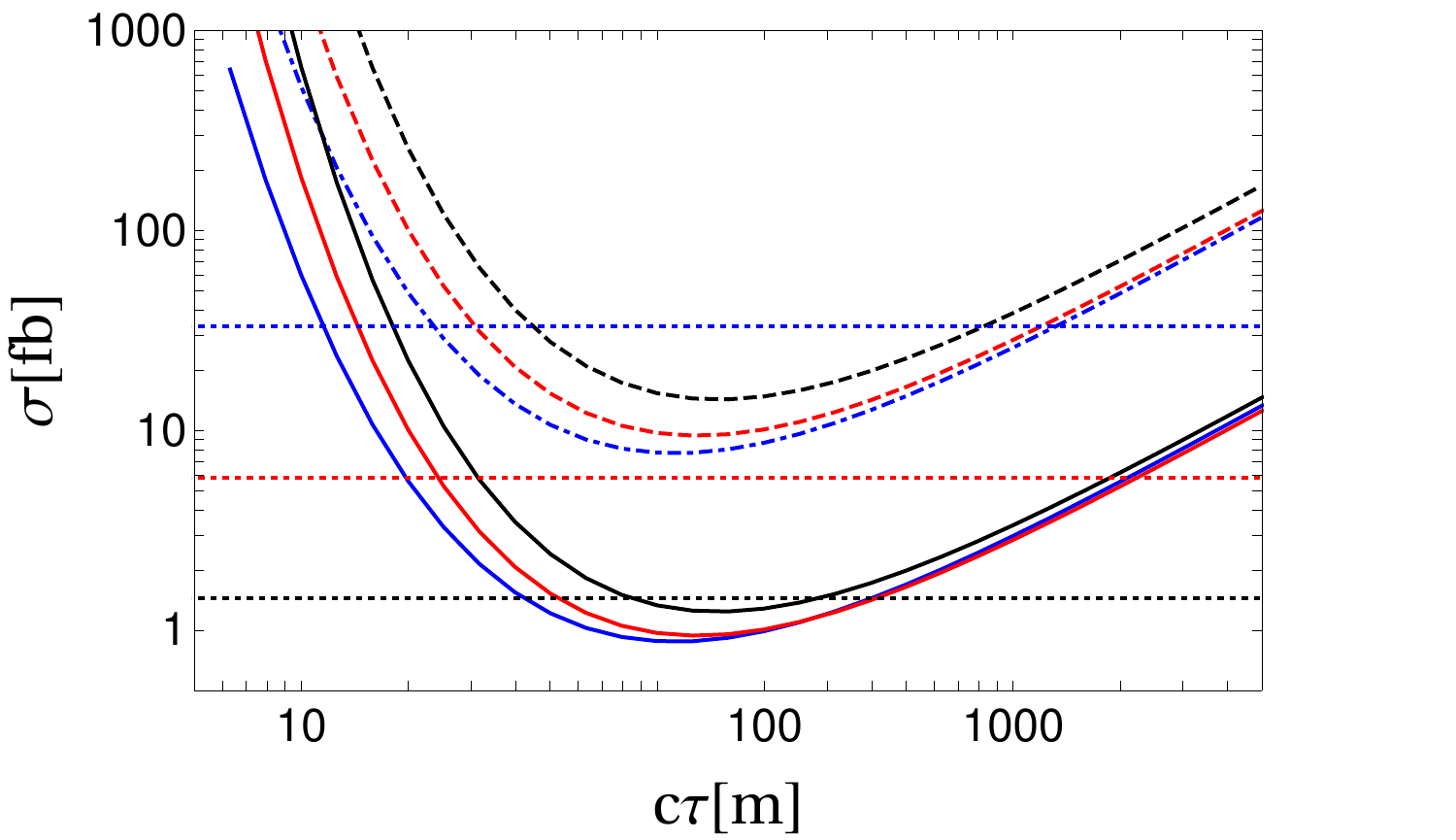}
	\caption{Discovery reach as a function of the proper LLNP decay-length $c\tau$ assuming LLNP production via Standard Model Higgs decay (left plot) or via Drell-Yan (right plot), for the same LLNP masses as in Fig.~\ref{fig:Hist_T}. The solid line correspond to an optimistic scenario where all the background can be removed, while the dashed line, to a conservative background model (see text for details). 
	We also show in the left panel the Higgs decay branching ratio into two LLNPs leading to the corresponding LLNP production cross section, and in the right panel, as dotted lines, the LLNP production cross section via Drell-Yan for the corresponding LLNP masses, for the specific case of the Higgsino as LLNP. }
	\label{fig:reach}
\end{figure}

MATHUSLA aims to be a zero-background detector. If the design goal is achieved, the observation of 4 events will be sufficient to ensure a $3\sigma $ detection. The probability to observe an event depends on the geometric coverage and the decay probability, {\it i.e.} the lifetime in the detector frame. Here and in the following we define an observable event as follows: we require that the LLNP  decays inside the MATHUSLA volume and that both final state leptons pass through the tracker layers on top of the volume (we neglect in our analysis the possibility that a lepton traverses the trackers escaping detection). In addition, we require that the opening angle of the lepton pair is $\geq 1\degree $, to ensure that the individual leptons are clearly separated and that the displaced vertex can be reconstructed to good precision. We simulate the production and decay of the LLNPs with CalcHEP~\cite{Belyaev:2012qa} and select events which fulfill all geometric requirements. The discovery reach, defined by the cross section leading to 4 observable events, is shown  in figure~\ref{fig:reach} both for scenario A (left panel) and for scenario B (right panel). The left panel also shows, for  reference, the Higgs decay branching ratio into two LLNPs leading to the corresponding LLNP production cross section~\cite{Higgsxsec,Dittmaier:2011ti}, and the right panel, the Drell-Yan production cross section for the specific example of the Higgsino as LLNP~\cite{susyxsec,Fuks:2012qx,Fuks:2013vua}, in  both  cases at the LHC running at $\sqrt{s}=14$ TeV.

As can be seen, MATHUSLA is sensitive to $\sigma\gtrsim 1 \,{\rm fb}$ both for scenario A and for scenario B, with a maximum sensitivity when the decay-length  is $c\tau\sim 100$~m. Even for a decay length of $5000$ m a production cross section $\sigma \gtrsim 10 \,\mbox{fb}$  can be probed. Note that the discovery reach is comparable to the one expected for scenarios with two-body LLNP decays \cite{Chou:2016lxi}, although one observes a slight degradation of the expected sensitivity as the LLNP mass increases, due to the different kinematics of the two- and the three-body decays.

The discovery reach could be affected by the existence of backgrounds. Possible sources of background in MATHUSLA are 
cosmic ray muons, high energy muons produced at the LHC and atmospheric neutrinos  \citep{Chou:2016lxi}. Muon events can be efficiently rejected using timing cuts (as most atmospheric muons are down-going) and using a scintillator veto layer surrounding the detector. On the other hand, background rejection of neutrino-induced events is more involved. 
 About $60$ neutrino events with a proton in the final state are expected per year \cite{Chou:2016lxi}. Most these events will only contain non-relativistic protons but $\approx 10\%$ of them are expected to have a fast moving proton in the final state \cite{DavidCurtin}. Consequently, one expects that six events per year will feature a fast moving proton and a charged lepton in the final state.     
 These events will then produce two tracks in the detector and will mimic a signal event. It has been argued that this source of background could be suppressed by requiring that the reconstructed LLNP trajectory points towards the LHC interaction point. However, this geometric veto implicitly assumes that the LLNP decays into two visible particles (producing the tracks) and cannot be straightforwardly applied to scenarios where the LLNP decays into three particles, one of them being invisible. 

\begin{figure}[t]
	\centering
	\includegraphics[width=0.45\textwidth]{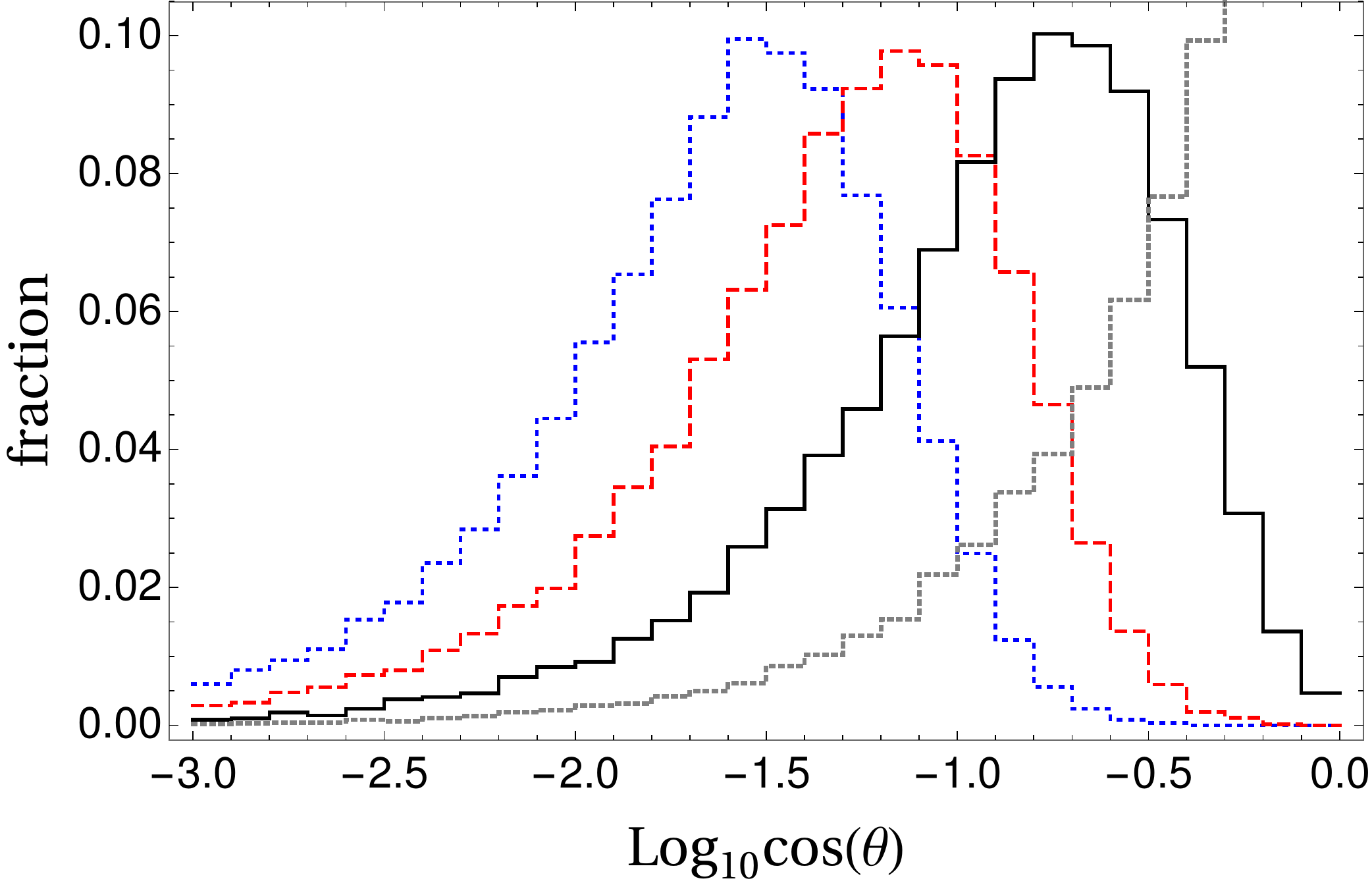}
	\includegraphics[width=0.45\textwidth]{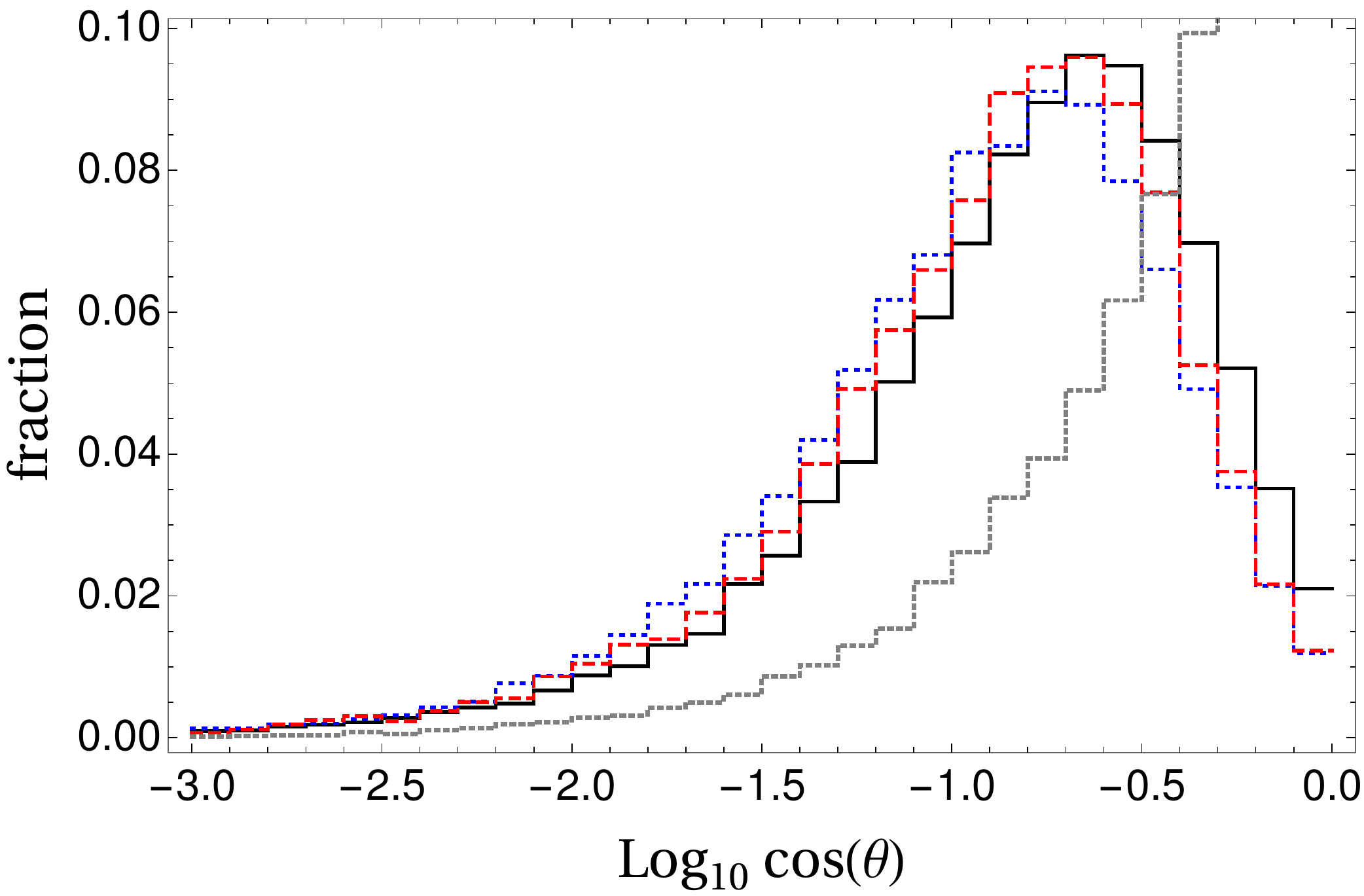}
	\caption{Same as Fig.~\ref{fig:Hist_T}, but as a function of the cosine of the angle between the LLNP direction and the perpendicular to the decay plane spanned by the two tracks. The expected distribution from atmospheric neutrino 
	induced events is indicated by the gray dashed line.
	} \label{fig:Hist_cos}
\end{figure}

A detailed analysis of the backgrounds in MATHUSLA is beyond the scope of this paper.\footnote{ We note in particular that the $T$-distribution could be used to reject potential backgrounds in a search for new physics from two-body decays.} However, one can estimate the impact of the neutrino background on the MATHUSLA discovery reach for three-body decays by noting that the flux of atmospheric neutrinos is isotropic. Therefore, there is no preferred incoming direction and, correspondingly, the angle between the vector joining the LHC detector with the point where the two tracks intersect, and any vector perpendicular to the plane formed by the two track directions, also follows an isotropic distribution. We show in Fig.~\ref{fig:Hist_cos} the distribution in the angle between the LLNP direction and the vector perpendicular to the plane spanned by the two tracks, $\cos\theta$, for representative choices of the LLNP mass, assuming scenario A (left panel) or scenario B (right panel); the atmospheric neutrino induced events are shown as gray dashed line. It is apparent from the figure that background events and signal events have a different $\cos\theta$-distribution, thus allowing to subtract background events. Let us note that the $T$-parameter depends not only on $\cos\theta$ but also on the opening angle between the observed tracks, $\sin\phi$ in Eq.~(\ref{eq:T-3body}). Therefore, we expect that the $T$-distribution would allow an even stronger discrimination between background and signal events. In this paper, and since a detailed understanding of the $\sin\phi$ dependence of the neutrino-induced track events is still lacking, we will use the $\cos\theta$ distribution to estimate a conservative discovery reach of three-body decays in MATHUSLA.

To determine the discovery reach we employ the test statistics on the $\cos\theta$ distribution. The 
binned Poisson likelihood  $L$ is given by~\cite{Tanabashi:2018oca}:
\begin{align}
-2 \log L(x)=2 \sum_i \left[ R_i(x) - N_i + N_i  \log \frac{N_i}{R_i (x)} \right]\;,
\end{align}
where $i$ runs over the number of bins, $N_i$ denotes  the number of events observed in bin $i$ while $R_i(x)$ is the expected number of signal and background events as a function of the theory parameters $x$. We perform 2500 pseudo-experiments and follow the approach described in \cite{Cowan:2010js} to determine the lowest cross-section leading to a $3\sigma$ discovery in $90\%$ of the pseudo-experiments. The result is shown in Fig.~\ref{fig:reach} as dotted lines for scenario A (left plot) and for scenario B (right plot). As expected, the sensitivity to new physics gets reduced when including the background. Yet, even with our very conservative assumptions for the background modeling, the discovery cross section increases by at most a factor $\approx3$ for light LLNPs, and by a factor $\approx 10$ for heavy LLNPs (note that heavier LLNPs move more slowly, and correspondingly the decay products are emitted widely separated, resembling more and more a background event). 

Let us stress that our assumptions for the background modeling are very conservative and that not many neutrino-induced events are expected to have the same characteristics as a three-body LLNP. Given the large LLNP production rate in Higgs decays or in Drell-Yan expected at the HL-LHC, the forecast number of signal events can be potentially huge, thus opening the possibility of disentangling the fundamental parameters of the BSM Lagrangian from observations. We will address this issue in the next section.

\section{Reconstruction of model parameters}
\label{sec:reconstruction}

\begin{figure}[t]
\centering
\includegraphics[width=0.7\textwidth]{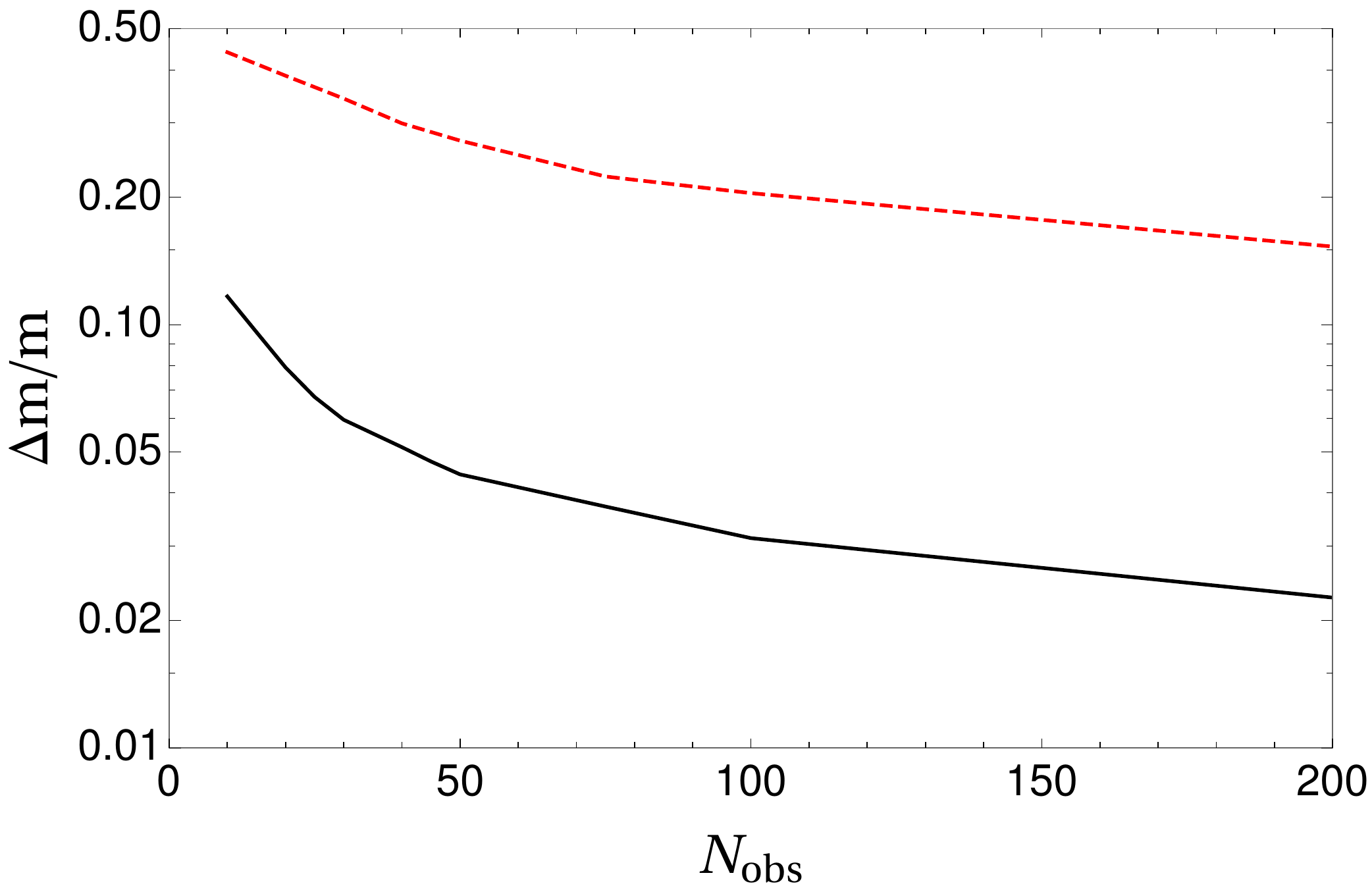}
\caption{Relative precision of the measured mass as a function of the observed number of events for a 25 GeV LLNP produced via Standard Model Higgs decays (black, solid) and a $600$ GeV LLNP produced via Drell-Yan (red, dashed).  \label{fig:massprecision}}
\end{figure}

In this section we will study the potential of the MATHUSLA detector to extract information about the underlying model parameters for the scenarios A and B.  
We conduct 1000 pseudo-experiments to reconstruct the best fit mass employing a maximum likelihood method.
We impose the full geometric cuts of the MATHUSLA detector with the design parameters described in Sec.~\ref{sec:23} and we require that the opening angle between the two leptons is larger than $1^\circ$, to ensure that the directions of both leptons and the position of the displaced vertex are measured with high accuracy. We also assume an optimistic case when all backgrounds can be removed.

In Fig.~\ref{fig:massprecision} we show the expected relative error in the determination of the LLNP mass as function of the number of observed events for one representative case in each of the scenarios of LLNP production under consideration. In the case of a $m_{\rm LLNP}=25$ GeV produced through the Higgs portal, a $10\%$ accuracy determination of the mass can be achieved with  less than 20 events, improving up to a $3\%$ accuracy if $200$ events are observed. MATHUSLA will then allow a fairly good determination of the LLNP mass provided the Higgs decay branching fraction into two LLNPs is $\gtrsim 10^{-4}$. For $m_{\rm LLNP}=600$ GeV produced by the Drell-Yan process, the mass reconstruction is poorer, as can be anticipated from the mild dependence of the $T$-distribution on the LLNP mass, and even 200 events would not allow a mass determination with an accuracy better than a $10\%$. Given that the production cross section of a 600 GeV Higgsino is just $5.8~\mbox{fb}$ \cite{susyxsec} an accurate determination of the mass based on the $T$ parameter does not seem feasible, unless additional production channels contribute substantially to Higgsino production at the LHC.  

 Finally, we would like to stress that this analysis is far from being exhaustive and should be regarded only as a first step towards the exploration of three-body decays in MATHUSLA. Generalizations to other kinematics of the decays, {\it i.e.} on-shell resonances in the three-body decay or a non-negligible $m_{\rm NSP}$ mass, may require a more elaborated treatment, as the mapping between the angular observables and the theory parameters becomes in this case more complicated. We also expect that the inclusion of more observables in the analysis will lead to a better parameter reconstruction and possibly to an eventual identification of the underlying model. 
For instance, a coordinated effort between the surface and the underground detectors may lead to the  observation of large amounts of missing transverse momentum at ATLAS/CMS and the correlated observation of track events in MATHUSLA, thus allowing to further suppress backgrounds and to tag the LLNP from the production point to the decay point. In particular, this would allow to determine the LLNP lifetime in the laboratory frame, and from the distance traveled, the LLNP velocity. Furthermore, the LLNP momentum could be determined from the missing transverse momentum (putatively carried by the LLNP) and from the angle between the LLNP direction and the LHC beam axis. In this way the LLNP mass could be reconstructed from the momentum and the velocity.


\section{Conclusions}

MATHUSLA is a proposed large-volume surface detector installed near ATLAS or CMS with a high capability of detecting two-track events and in reconstructing the track directions. These events can originate in the decay of a LLNP into two visible particles, or in the decay of a LLNP into three particles (two visible and one invisible) or from the charged current interaction of a neutron in the detector with an energetic neutrino of atmospheric origin. Previous works have shown that MATHUSLA offers excellent prospects to discriminate over the expected backgrounds signals of new physics when the LLNP decays into two Standard Model particles and the decay length is between $\sim 1$ m and $\sim 10^7$m. 

In this paper we have focused on scenarios where the LLNP decays into two visible and one invisible particle, possibly the dark matter particle. This scenario arises, {\it e.g.} in models with low scale supersymmetry breaking (where the lightest neutralino decays into a lepton-antilepton pair and a gravitino), $R$-parity non-conservation (where the lightest neutralino decays into a lepton-antilepton pair and a neutrino), feebly interacting dark matter, or sterile neutrinos. 

We have proposed the triple product of the two track directions and the direction between the decay point and the LHC production point as an efficient measure to discriminate between two-body, three-body and background events. We have analyzed two benchmark scenarios where the LLNP is produced via Drell-Yan and where it is produced via Standard Model Higgs decay, and we have shown that, in a zero-background experiment, LLNP three-body decays could be detected and discriminated from two-body decays and from background provided the decay length is sufficiently short (smaller than a few tens of kilometers). We have also estimated that in the presence of backgrounds the discovery reach is reduced, but by a factor smaller than $\approx 3$ for light LLNPs and smaller than $\approx 10$ for heavy LLNPs. Finally, we have briefly addressed the prospects of reconstructing the LLNP mass with MATHUSLA. 

Our results encourage more detailed analyses of the backgrounds in MATHUSLA, of the complementarity with possible correlated signals of new physics at ATLAS or CMS, and applications to concrete particle physics models.

\section*{Acknowledgments}

The work of AI has been partially supported by the DFG cluster of excellence EXC 153 ``Origin and Structure of the Universe'' and by the Collaborative Research Center SFB1258.  SV thanks G. Arcadi for helpful discussions. EM thanks the Department of Physics and Astronomy of Aarhus University for the hospitality during the completion of this paper.

\bibliographystyle{JHEP_improved}
\bibliography{refs}

\end{document}